\documentclass[preprint,prb,superscriptaddress]{revtex4}

\usepackage{graphicx}
\usepackage{dcolumn}
\usepackage{bm}

\begin{document}

\title{Giant magnetocaloric effect near room temperature in the off-stoichiometric Mn-Co-Ge alloy}

\author{V. K. Sharma}
\email{vishnusharma@rrcat.gov.in} 
\author{M. A. Manekar}
\affiliation{Magnetic and Superconducting Materials Section \\ Raja Ramanna Centre for Advanced Technology, Indore 452013, India}
\author{Himanshu Srivastava}
\affiliation{Indus Synchrotrons Utilization Division \\ Raja Ramanna Centre for Advanced Technology, Indore 452013, India}
\author{S. B. Roy}
\affiliation{Magnetic and Superconducting Materials Section \\ Raja Ramanna Centre for Advanced Technology, Indore 452013, India}

\date{\today}

\begin{abstract}
We report a giant magnetocaloric effect near room temperature in an off-stoichiometric Mn-Co-Ge alloy, across  the magnetostructural transition. The isothermal entropy change accompanying this transition has a peak value of nearly 40 J/kg-K near 297 K and a refrigerant capacity of 270 J/kg with the hot end at 302.5 K and cold end at 293.5 K. We also present an experimental protocol to avoid spurious peaks in the magnetocaloric effect across a sharp first order magnetostructural transition, not confined to Mn-Co-Ge alone, where metastability during the transition could influence the measured magnetization and thus the estimated entropy change. The estimated entropy change in the present off-stoichiometric Mn-Co-Ge alloy is possibly the highest reported value near room temperature in undoped Mn-Co-Ge alloys and underlines the potential of the alloy for technological applications in room temperature magnetic refrigeration.   
\end{abstract}

\pacs{75.30.Sg}
\keywords{Magnetocaloric effect, Magneto-structural transition}
\maketitle

The magnetocaloric effect (MCE) based cooling technology offers superior alternative to  the present day gas compression based cooling technology as it is more efficient as well as environment friendly.\cite{Tishin} Generally, MCE arises from the effect of adiabatic application/removal of magnetic field on the spin entropy of the system. In materials undergoing first order magnetostructural transition the structural entropy also contributes to this MCE,\cite{Pecharsky} thus giving rise to very large total entropy changes. Such materials are potential candidates for MCE based refrigeration technology and various Gd-Ge,\cite{Pecharsky} Mn-As,\cite{Gama} Ni-Mn-In\cite{Sharma} and Fe-Rh\cite{Manekar, Annaorazov} based alloys exhibiting first order magnetostructural transition in different temperature regimes have been investigated intensively in this regard.

Recently, the Mn-Co-Ge based alloys have drawn worldwide attention for their large MCE.\cite{Fang, Bao, Caron, Samanta, Trung2} The stoichiometric composition of ternary alloy MnCoGe undergoes a martensitic transition around 650 K from high temperature Ni$_2$In-type hexagonal phase (space group P63/mmc) to low temperature TiNiSi-type orthorhombic phase (space group Pnma).\cite{Trung2} The hexagonal phase can be obtained in a metastable form at room temperature by rapid cooling from high temperatures. Both the hexagonal and orthorhombic phases are ferromagnetic with Curie temperature ($T_C$) around 275 K and 345 K respectively.\cite{Trung1} The structural and magnetic transitions in the stoichiometric MnCoGe are thus entirely decoupled. The orthorhombic phase has a higher saturation magnetic moment.\cite{Trung1} The structural transition temperature in the alloy is reported to shift with change in stoichiometry,\cite{Fang} introduction of vacancies\cite{Liu} or interstitials,\cite{Trung1} doping with fourth element,\cite{Bao, Caron, Samanta, Trung2, Wang} and/or under external pressure.\cite{Niziol} These control parameters can tune the structural transition so as to lie between the Curie temperatures of the orthorhombic and hexagonal phases, and thus leading to the coupling of magnetic and structural transitions. Recent works report the tuning of this magnetostructural transition to near room temperature with careful doping. These doped Mn-Co-Ge alloy compositions exhibit a transition from paramagnetic hexagonal phase to ferromagnetic orthorhombic phase, and are the focus of MCE related research activities to attain MCE near ambient temperature.\cite{Fang, Bao, Caron, Samanta, Trung2} In this work, we have studied an off-stoichiometric Mn-Co-Ge alloy, without any doping, that undergoes a magnetostructural transition around ambient temperature and exhibits a large MCE amounting to an isothermal entropy change of nearly 40 J/kg-K. This is possibly the highest reported MCE around room temperature in undoped Mn-Co-Ge alloys. Thus the alloy is a potential candidate for technological application for magnetic refrigeration near room temperature. We also report the precautions to be taken during the experimental determination of the entropy change across a very sharp first order magnetostructural transition, which should hopefully establish a standard protocol for other material systems showing giant MCE with an underlying first order transition.

The polycrystalline Mn-Co-Ge sample was prepared by melting the required amount of pure elements in an arc melting furnace under inert argon gas atmosphere. The as cast sample was sealed in a quartz tube in an argon atmosphere and annealed at 1123 K for 5 days for homogenization, followed by controlled cooling at 5 K/hour to room temperature. The resulting sample was characterized with x-ray fluorescence (XRF) for composition and x-ray diffraction (XRD) for structural analysis. XRF measurements were performed on the microprobe-XRF beamline (BL-16) of the Indus-2 synchrotron facility using an excitation energy of $\sim$12 keV.\cite{Tiwari} A Vortex\textsuperscript{TM} Si drift detector (SDD) was used to measure the characteristic fluorescence emission from the MnCoGe sample. Quantification was done using a homemade computer program based on the fundamental parameter method.\cite{Rene} The composition was found to be Mn$_{34.5}$Co$_{33.1}$Ge$_{32.4}$ within the error bar of 1\%. However, in the following discussion the sample will be referred as MnCoGe. XRD measurement on the powdered sample was performed with a commercial diffractometer (D8 Discover, Bruker AXS) using Cu K$_{\alpha}$ radiation. The magnetization ($M$) measurements as functions of temperature ($T$) and magnetic field ($H$) were performed in a commercial SQUID magnetometer (MPMS-7XL, Quantum Design, USA). Three different standard protocols were adopted for the $M$ versus $T$ measurements in a constant applied magnetic field: zero field cooled (ZFC) warming, field cooled cooling (FCC) and field cooled warming (FCW).  Isothermal $M(H)$ curves were measured starting from an initial state prepared by unidirectional cooling from 350 K to the target temperature in zero applied field.

\begin{figure}
\includegraphics{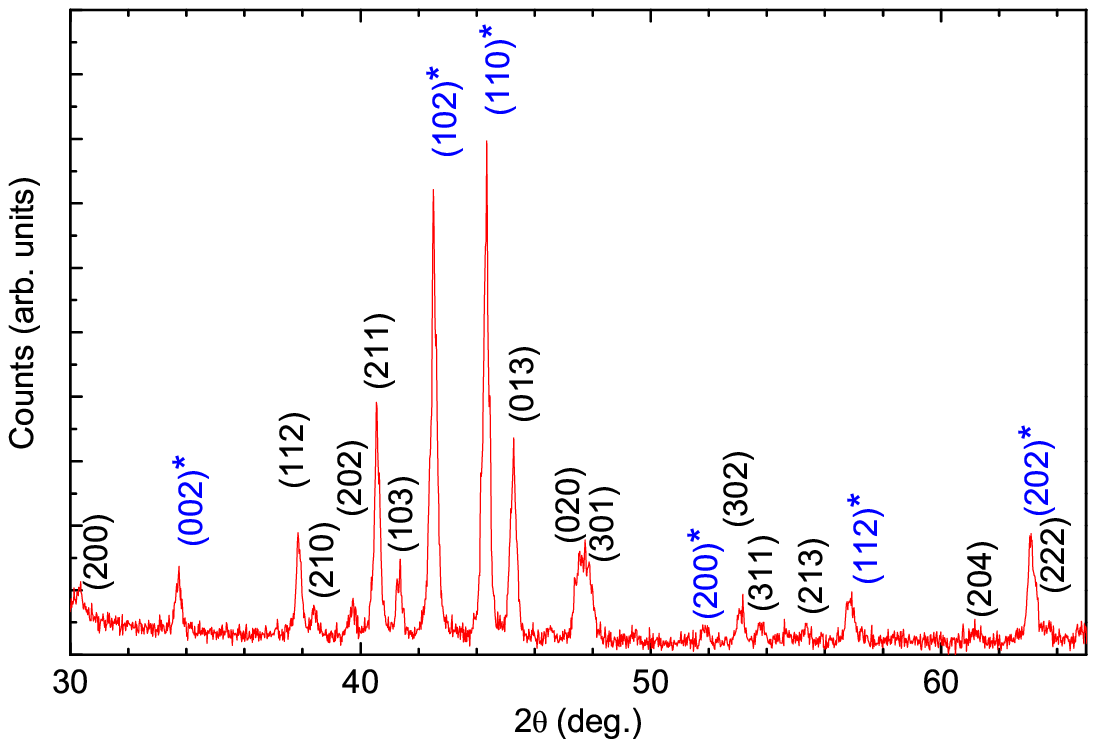}
\caption{\label{xrd}X-ray diffraction pattern of MnCoGe at $T$=295 K. Peaks corresponding to the hexagonal phase and the orthorhombic phase are marked with and without `$\star$' respectively}
\end{figure}

Room temperature ($\sim$295K) XRD pattern of MnCoGe is presented in fig. \ref{xrd}. The complete pattern could not be indexed to a single hexagonal or a single orthorhombic structure. This suggests the presence of both the phases in the alloy at room temperature. The lattice parameters of orthorhombic phase are found to be a=5.91\AA, b=3.81\AA\ and c=7.03\AA. The lattice parameters of hexagonal phase are a=4.07\AA\ and c=5.30\AA. As per the relationship between lattice parameters of the orthorhombic and hexagonal phases in Mn-Co-Ge alloys, two hexagonal unit cells transform to one orthorhombic unit cell.\cite{Johnson} This gives an estimation of volume change from the hexagonal to orthorhombic phase as nearly 4\% . We have observed that the sample breaks into smaller pieces after repeated temperature cycling through the structural transition regime. Probably such a large volume change is responsible for the shattering of the sample across the transition.

\begin{figure}
\includegraphics{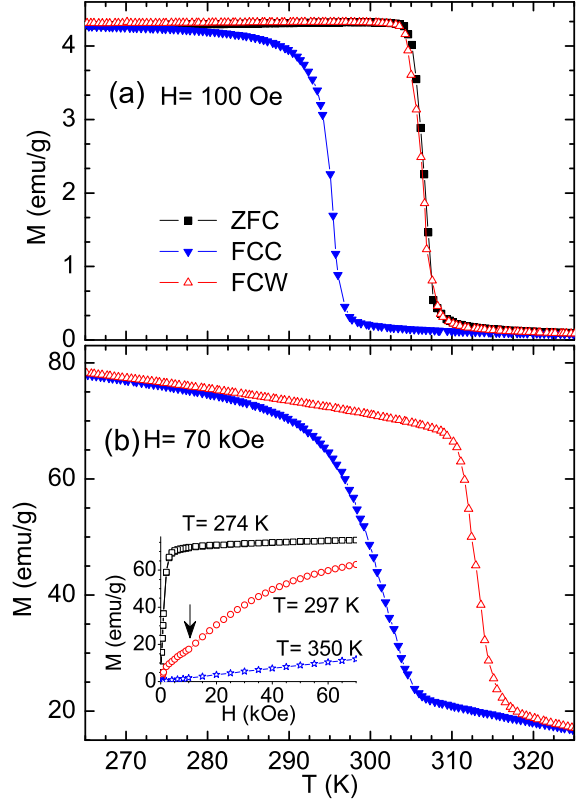}
\caption{\label{MT}Temperature dependence of magnetization of MnCoGe in applied magnetic fields of (a) 100 Oe and (b) 70 kOe.  Inset to (b) shows the isothermal field dependence of magnetization measured at representative temperatures in orthorhombic phase (274K), hexagonal phase (320K) and across the hexagonal-orthorhombic transition region (297K). Arrow marks the onset of the metamagnetic transition. }
\end{figure}

The temperature dependence of magnetization of MnCoGe in $H$ = 100 Oe is presented in fig. \ref{MT} (a). Along the FCC curve, the sharp rise in magnetization at nearly 298 K indicates a transition from the paramagnetic (low magnetic moment) hexagonal phase to the ferromagnetic (higher magnetic moment) orthorhombic phase. This transition is accompanied by a thermal hysteresis, which is a typical signature of a first order transition.\cite{Chaikin}  Existing literature\cite{Trung2} establishes that the phase at lower temperature ($T<$275K) is the orthorhombic phase and the phase at higher temperature ($T>$315K) is the hexagonal phase. The onset of the hexagonal to orthorhombic transition during cooling occurs at a lower temperature ($\sim$298K) than the onset of the orthorhombic to hexagonal transition during heating ($\sim$304K). This relationship between the onset temperatures of the phase transformations during cooling and heating indicates that the transition is only moderately influenced by disorder\cite{Chattopadhyay} and there is only a narrow temperature range of coexistence of both the phases. At 295 K along the low field FCC curve (fig. \ref{MT}(a)), the transition is not complete, so we expect coexisting hexagonal and orthorhombic phases at this temperature. This explains the XRD pattern presented in fig. \ref{xrd} where peaks corresponding to both the hexagonal phase as well as the orthorhombic phase are observed. In higher magnetic field the transition is shifted to higher temperature as seen from $M(T)$ curves in $H$=70 kOe  presented in fig. \ref{MT}(b). This indicates to a possibility of field induced transition from hexagonal phase to orthorhombic phase at constant temperature.

To explore the possibility of field induced transition in the alloy we have measured the isothermal $M(H)$ curves. The inset of fig. \ref{MT}(b) shows the $M(H)$ curves at three representative temperatures. Isothermal $M(H)$ curves measured at temperature 274 K, corresponding to complete  orthorhombic  phase, is typical of a ferromagnet. On the other hand, the isothermal $M(H)$ curve measured in hexagonal phase at 350 K suggests a paramagnetic nature of the phase at this temperature. These observations are in congruence with the existing literature.\cite{Trung2} The isothermal $M(H)$ curve around 297 K depicts a substantially different nature as compared to $M(H)$ curves at 274 K and 350 K. The lack of tendency to saturation till 10 kOe in $M(H)$ at 297 K is related to the coexisting paramagnetic hexagonal phase and ferromagnetic orthorhombic phase. At around 10 kOe field, the $M(H)$ curve at 297 K shows a metamagnetic character. This suggests a field induced transition from low magnetization hexagonal phase to the high magnetization orthorhombic phase indicating a strong magnetostructural coupling in the alloy. Such a coupling could lead to large entropy change comprising of spin and structural degrees across the transition, and is expected to give rise to large MCE in the alloy. To investigate the MCE in MnCoGe alloy, we have estimated the isothermal magnetic entropy change ($\Delta S_M$) around room temperature by using the Maxwell relation, 

\begin{equation}
\Delta S_M (T,H) = \int\limits_{0}^{H}\Bigg[\frac{\partial M(T)}{\partial T}\Bigg]_H dH.
\end{equation}

\begin{figure}
\includegraphics{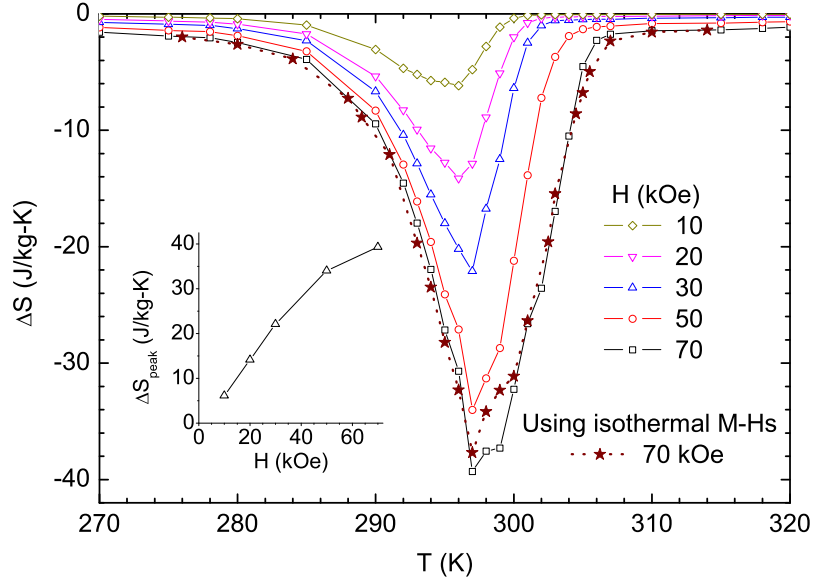}
\caption{\label{mce}Isothermal entropy change in MnCoGe sample for different magnetic field excursions estimated from the $M(T)$ curves in constant magnetic fields. Entropy change in 70 kOe field as estimated from isothermal $M(H)$ curves is also shown for comparison. Magnetic field dependence of peak value of isothermal entropy change is shown in the inset.}
\end{figure}

Generally, $\Delta S_M$ is estimated from the isothermal $M(H)$ curves measured at discrete field intervals by numerically integrating Maxwell’s relation.\cite{McMichael} However, it is seen that $\Delta S_M$ estimation from $M(H)$ curves across a first order transition could be erroneous due to a change in the ferromagnetic and paramagnetic phase fractions (in zero field) as a function of temperature.\cite{Caron2, Tocado,Cui} Here we have estimated $\Delta S_M(T)$ from the numerical integration of Maxwell’s relation using both the FCC $M(T)$ curves\cite{Sharma2} in various fields and isothermal $M(H)$ curves at various temperatures. For comparing the entropy change estimated from two protocols, the initial state of the sample should be the same. The entropy change is estimated across the hexagonal to orthorhombic transition which occurs either by reducing the temperature or by the application of magnetic field at high temperature.  Thus the  FCC curves instead of the ZFC (warming) curves have to be used. The isothermal $M(H)$ cycle at a particular temperature should be started after cooling the sample from high temperature in zero field. The comparison of $\Delta S_M$ estimated from the FCC $M(T)$ curves in various fields and isothermal $M(H)$ curves for a field change of 70 kOe is presented in fig. \ref{mce}.  The estimated $\Delta S_M$ shows a peak around 297 K with a value of 6.2 J/kg-K for a field excursion of 10 kOe, which increases to 39.3 J/kg-K for a field change of 70 kOe (see inset to fig. \ref{mce}). This $\Delta S_M$ observed in MnCoGe is much larger than the $\Delta S_M$ $\sim$10 J/kg-K observed in Gd around 295 K for a field excursion of 70 kOe.\cite{Benford}  $\Delta S_M$ = 34 J/kg-K observed in the present alloy for a field change of 50 kOe is also larger than 19 J/kg-K reported in Gd$_5$Si$_2$Ge$_2$ alloy for the same field.\cite{Pecharsky2} When we compare the performance of the present MnCoGe sample with other materials of the MnCoGe family, we find that for a field change of 10 kOe, $\Delta S_M$ is comparable to that of 6.4 J/kg-K reported around 330 K in MnCo$_{0.95}$Ge$_{1.14}$.\cite{Fang} The observed value of 34 J/kg-K in the present sample for a field excursion of 50 kOe is larger than the highest reported value of 28.5 J/kg-K around 320 K in Mn$_{0.96}$Cr$_{0.04}$CoGe\cite{Trung2} and also larger than 17.2 J/kg-K achieved around 300 K in Mn$_{0.92}$Fe$_{0.08}$CoGe\cite{Wang} This entropy change is similar to 35 J/kg-K in MnCoGe$_{0.8}$Al$_{0.2}$ at 322 K,\cite{Bao} smaller than 53 J/kg-K observed around 321K in Mn$_{0.92}$Cu$_{0.08}$CoGe\cite{Samanta} and 47 J/kg-K around 287 K in MnCoGeB$_{0.2}$.\cite{Trung1} However, some of the $\Delta S_M$ values reported in literature have been estimated from isothermal $M(H)$ curves which can result in higher estimated values.\cite{Tocado, Cui} On the other hand, the method of $\Delta S_M$ estimation from isofield $M(T)$ curves employed here is more reliable.\cite{Tocado} We note that the $\Delta S_M$ of 40 J/kg-K and 34 J/kg-K obtained in present MnCoGe alloy for field changes of 70 kOe and 50 kOe respectively, are the highest reported $\Delta S_M$ near room temperature in undoped Mn-Co-Ge alloys. The refrigerant capacity (RC) which is a measure of the heat flow from cold end to hot end in an ideal refrigeration cycle, has been determined using the method of Gschneidner et al.\cite{Gschneidner} For the present alloy, RC is estimated to be 270 J/kg for a field change of 70 kOe with the cold and hot ends at 293.5 K and 302.5 K respectively. Thus the present MnCoGe has potential for room temperature magnetic cooling applications as the peak in the entropy change occurs at 297 K.

It is reported by various authors that the $\Delta S_M$ curve estimated from isothermal $M(H)$ curves in many alloy systems contains spurious spikes, and such features were interpreted to be arising due to the change in ferromagnetic and paramagnetic phase fractions with increase in temperature across a first order transition.\cite{Caron2, Tocado,Cui} In our case, we have also noticed many a times that the estimation of $\Delta S_M$ from isothermal $M(H)$ curves gives quite unreliable values which are not reproducible. This happens despite the fact that after each isothermal $M(H)$ measurement the sample was heated to high temperature (350K) which is well above the first order transition regime and then cooled in zero field to the target temperature of next $M(H)$ measurement. This $M(H)$ measurement protocol is identical to the `loop process' recommended by Caron et al.\cite{Caron2} for $M(H)$ measurements.  We find that the lack of reproducibility in $\Delta S_M$ can be avoided if all the $M(H)$ curves are measured one after the other in a single experimental cycle without pausing between measurements or switching off the instrument, while following the loop process. Thus After each isothermal measurement, the sample is heated back to the high temperature reversible state (350K) and then cooled in zero field till the target temperature value and the instrument is kept ON till the all the $M(H)$ curves are measured in the desired temperature range. This ensures a similar temperature environment in the sample chamber for all the measurements and therefore an equal rate of temperature variation across the metastable region to reach the initial temperature point for $M(H)$ measurement, thereby minimizing the effect of metastability (and thus finite relaxation) across the transition. By following this particular method we observe that the entropy change estimated from $M(T)$ and $M(H)$ curves are in close agreement to each other with no spurious peaks in the $\Delta S_M$ curve. Since the measurements in a SQUID magnetometer are quite slow, the entire experimental cycle needs a few days of continuous time for obtaining reproducible results.

In conclusion, the off-stoichiometric Mn-Co-Ge alloy investigated in this study shows a field induced transition from paramagnetic (hexagonal) phase to ferromagnetic (orthorhombic) phase around room temperature. This gives rise to a large magnetocaloric effect in the alloy at ambient temperatures. The isothermal entropy changes of around 34 J/kg-K and 40 J/kg-K for field change of 50 kOe and 70 kOe respectively, observed in this alloy, are the largest reported values around room temperature in undoped Mn-Co-Ge alloys. Refrigerant capacity for a field change of 70 kOe is 270 J/kg with the hot and cold ends at 302.5 K and 293.5 K respectively. The alloy system could be useful for technological applications as it breaks up into very small pieces within a few field-temperature cycles due to a large volume change across the transition. This can naturally offer a large surface area for heat exchange when the alloy will be used in magnetic cooling applications. We have also devised an experimental protocol to avoid the spurious peaks in estimation of entropy change from isothermal $M(H)$ curves by initiating the measurement cycle after following the same rate of temperature variation for all the measurements.

\begin{acknowledgments}

The authors acknowledge the help of R. K. Meena in sample preparation, M. Murugan in sealing of sample, Anil Chouhan in heat treatment of sample, Ajit Kumar Singh and Khooha Ajay Laxmandas in XRF measurements and Dr. S. K. Rai for XRD measurements. The authors also acknowledge the help of Cryo-engineering and Cryo-module Development Section, RRCAT, Indore in providing cryogens.

\end{acknowledgments}

\newpage

\end{document}